\documentclass[conference]{IEEEtran}
\IEEEoverridecommandlockouts

\usepackage[normalem]{ulem}
\usepackage{cite}
\usepackage{amsmath,amssymb,amsfonts}
\usepackage{algorithmic}
\usepackage{graphicx}
\usepackage{textcomp}
\usepackage{xcolor}
\usepackage{multirow}
\usepackage{booktabs}
\usepackage{tikz}
\usepackage{pgf}
\usepackage{arydshln}
\usepackage{url}
\usepackage{siunitx}
\usepackage[normalem]{ulem}
\usepackage{hyperref}

\usetikzlibrary{arrows}
\usetikzlibrary{quotes}
\usetikzlibrary{arrows.meta}
\def\BibTeX{{\rm B\kern-.05em{\sc i\kern-.025em b}\kern-.08em
    T\kern-.1667em\lower.7ex\hbox{E}\kern-.125emX}}

\tikzstyle{specialblock} = [draw, ultra thick, fill=blue!20, rectangle, 
    minimum height=3em, minimum width=4em]
\tikzstyle{block} = [draw, fill=lightgray, rectangle, 
    minimum height=3em, minimum width=4em]
\tikzstyle{sum} = [draw, fill=white, circle, node distance=1cm]
\tikzstyle{prod}   = [circle, minimum width=8pt, draw, inner sep=0pt, path picture={\draw (path picture bounding box.south east) -- (path picture bounding box.north west) (path picture bounding box.south west) -- (path picture bounding box.north east);}]
\tikzstyle{sumt}   = [circle, minimum width=8pt, draw, inner sep=0pt, path picture={\draw (path picture bounding box.east) -- (path picture bounding box.west) (path picture bounding box.south) -- (path picture bounding box.north);}]
\tikzstyle{input} = [coordinate]
\tikzstyle{output} = [coordinate]
\tikzstyle{pinstyle} = [pin edge={to-,thin,black}]
\tikzset{
tmp/.style  = {coordinate}, 
dot/.style = {circle, minimum size=#1,
              inner sep=0pt, outer sep=0pt},
dot/.default = 6pt 
}

\begin{document}

\title{LMAC-TD: Producing Time Domain Explanations for Audio Classifiers
\thanks{\textsuperscript{*}Both authors contributed equally to this research. For these authors, the order is alphabetical.
\\ 
\\ This work has been submitted to the IEEE for possible publication. Copyright may be transferred without notice, after which this version may no longer be accessible.}
 }

\author{
    \IEEEauthorblockN{
        Eleonora Mancini\textsuperscript{* 1}, 
        Francesco Paissan\textsuperscript{* 2,3}, 
        Mirco Ravanelli\textsuperscript{3,4}, 
        Cem Subakan\textsuperscript{3,4,5}
    }
    \IEEEauthorblockA{
        \textsuperscript{1}DISI, University of Bologna, Italy 
        \textsuperscript{2}Fondazione Bruno Kessler, Italy 
        \textsuperscript{3}Mila-Québec AI Institute, Canada  \\
        \textsuperscript{4}Concordia University, Canada 
        \textsuperscript{5}Laval University, Canada
    } 
}

\maketitle

\begin{abstract}

Neural networks are typically black-boxes that remain opaque with regards to their decision mechanisms. Several works in the literature have proposed post-hoc explanation methods to alleviate this issue. This paper proposes LMAC-TD, a post-hoc explanation method that trains a decoder to produce explanations directly in the time domain. This methodology builds upon the foundation of L-MAC, Listenable Maps for Audio Classifiers, a method that produces faithful and listenable explanations. We incorporate SepFormer, a popular transformer-based time-domain source separation architecture. We show through a user study that LMAC-TD significantly improves the audio quality of the produced explanations while not sacrificing from faithfulness. 

\

\end{abstract}


\section{Introduction}

Black-box neural networks obtain impressive performance across various application domains, but their decision mechanisms typically remain opaque. Post-hoc explanation methods aim to alleviate this issue by highlighting the parts of the input that are deemed most influential for the network decisions \cite{Selvaraju_2019, smilkov2017smoothgrad, Chattopadhay_2018}. Post-hoc interpreters do not alter the original network architecture and operate \textit{post-hoc} on a pre-trained model. {Such methods} produce interpretations without constraining the network to interpretable designs~\cite{loiseau2022model, zinemanas2021interpretable, 9914699} or training strategies~\cite{flint} that can eventually harm performance.

In recent years, significant progress has been made in generating post-hoc listenable explanations. Notable works include SLIME \cite{DBLP:conf/ismir/MishraSD17} and AudioLIME \cite{haunschmid2020audiolime}. SLIME\cite{DBLP:conf/ismir/MishraSD17} segments the spectrogram into time-frequency regions similar to LIME's superpixels for images \cite{lime} and evaluates each region's importance. AudioLIME \cite{haunschmid2020audiolime} extracts sources from the audio and assigns a saliency score to each. L2I \cite{l2i} introduced a decoder to selectively activate an NMF dictionary based on the relevance of its components to the predicted class. L-MAC \cite{paissan24lmac} later simplified L2I's architecture by introducing a custom loss function to maximize the interpretation's faithfulness while removing the need for a pre-trained NMF dictionary. In LMAC-ZS~\cite{paissan2024listenablemapszeroshotaudio}, the authors adapt the loss function to work with zero-shot classifiers.
L-MAC generates more faithful and higher-quality explanations than its counterparts. However, it operates in the magnitude-STFT domain with a simple separation architecture that uses the input mixture's phase to reconstruct the waveform, thus causing artifacts. 

Also recently, deep learning-based source separation has achieved remarkable performance, reaching Signal-to-Noise Ratios above 20dB on benchmarks like WSJ0-2Mix \cite{wsj0-2mix} and Libri2Mix \cite{cosentino2020librimix}. Among the state-of-the-art models is SepFormer \cite{sepformer}. In this paper, we propose {enhancing the quality of L-MAC's explanations by} incorporating the SepFormer's MaskNet {in the interpreter}. That is, we generate explanations directly in the time domain, bypassing the need to use the input signal's phase as required by L-MAC's magnitude-STFT approach.

Our experimental results show that the proposed LMAC-TD, the time-domain version of the L-MAC framework, achieves comparable results in terms of the faithfulness metrics established in the original L-MAC paper \cite{paissan24lmac} while delivering improved audio quality as demonstrated by our user study.

\usetikzlibrary{arrows.meta}
\tikzstyle{dictsmall} = [draw, thick, fill=white!10, rectangle, 
    minimum height=1.0cm, minimum width=5cm] 
    \newcommand{\xshifts}{+4.7}
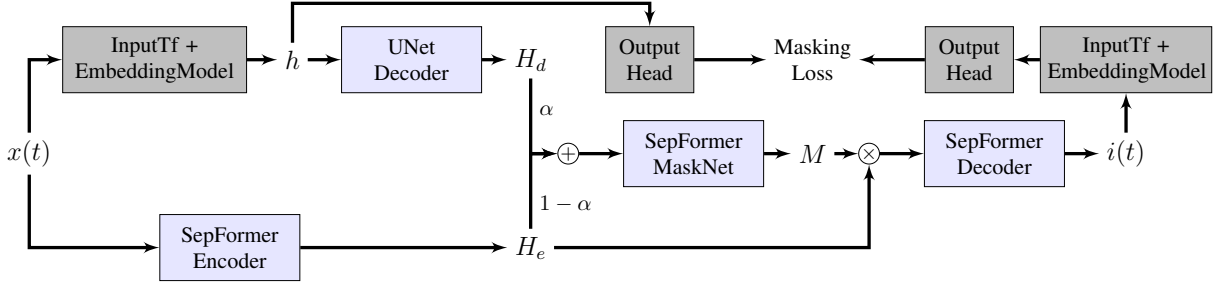
\begin{figure*}[t]
    \centering
    \resizebox{0.90\textwidth}{!}{
    \begin{tikzpicture}[auto, node distance=1.5cm,>=latex']
        \node [fill=none, xshift=-1cm] (input) {\large $x(t)$};
        \node [fill=none, right of=input] (ghst1) {};
        
        \node [block, above of=ghst1, xshift=0.5cm, text width=2.7cm, align=center] (tf) {InputTf + \\ EmbeddingModel};
        \node [fill=none, right of=tf, xshift=0.7cm] (h) {\large $h$};
        \node [block, right of=h, xshift=0.4cm, text width=2cm, align=center, fill=blue!10] (dec) {UNet \\ Decoder};
        \node [fill=none, above of=dec, yshift=-0.6cm, text width=2cm] (unetghst) {};
        \node [fill=none, right of=dec, xshift=0.4cm] (d) {\large $H_d$};
        \node [block, right of=d, xshift=0.4cm, yshift=.0cm, text width=1cm, align=center] (outhead1) {Output \\ Head};
        
        \node [block, below of=ghst1, xshift=1.7cm, text width=2cm, align=center, fill=blue!10] (sep_encoder) {SepFormer\\Encoder};
        \node [fill=none, below of=d, yshift=-1.5cm] (e) {\large $H_e$};
        \node [circle, draw=black, minimum size=0pt, inner sep=0pt, below of=d, xshift=0.6cm] (sum) {\small $+$};
        
        \node [block, right of=sum, text width=2cm, xshift=0.5cm, align=center, fill=blue!10] (masknet) {SepFormer\\MaskNet};
        \node [fill=none, right of=masknet, xshift=0.4cm] (m) {\large $M$};
        \node [circle, draw=black, minimum size=0pt, inner sep=0pt, right of=m, xshift=-0.6cm] (prod) {\small $\times$};
        \node [block, right of=prod, text width=2cm, xshift=0.5cm, align=center, fill=blue!10] (sep_dec) {SepFormer\\Decoder};
        \node [fill=none, right of=sep_dec, xshift=0.6cm] (int) {\large $i(t)$};
        \node [block, above of=int, text width=2.5cm, align=center] (cls) {InputTf +\\ EmbeddingModel};
        \node [block, left of=cls, text width=1cm, xshift=-1cm, align=center] (outhead2) {Output\\Head};
        \node [fill=none, left of=outhead2, xshift=-1cm, align=center] (loss) {Masking \\ Loss};

        \draw [->, line width=1.5pt] (input) |- (tf);
        \draw [->, line width=1.5pt] (input) |- (sep_encoder);
        \draw [->, line width=1.5pt] (tf) -- (h);
        \draw [->, line width=1.5pt] (h) -- (dec);
        \draw [->, line width=1.5pt] (dec) -- (d);
        \draw [->, line width=1.5pt] (sep_encoder) -- (e);
        \draw [->, line width=1.5pt] (d) |- node[right, yshift=0.7cm] {$\alpha$} (sum);
        \draw [->, line width=1.5pt] (e) |- node[right, yshift=-0.8cm] {$1 - \alpha$} (sum);
        \draw [->, line width=1.5pt] (sum) -- (masknet);
        \draw [->, line width=1.5pt] (masknet) -- (m);
        \draw [->, line width=1.5pt] (m) -- (prod);
        \draw [->, line width=1.5pt] (e) -| (prod);
        \draw [->, line width=1.5pt] (prod) -- (sep_dec);
        \draw [->, line width=1.5pt] (sep_dec) -- (int);
        \draw [->, line width=1.5pt] (int) -- (cls);
        \draw [->, line width=1.5pt] (cls) -- (outhead2);
        \draw [->, line width=1.5pt] (h) |- (unetghst.center) -| (outhead1);
        \draw [->, line width=1.5pt] (outhead1) -- (loss);
        \draw [->, line width=1.5pt] (outhead2) -- (loss);

    \end{tikzpicture}
    }
    \caption{The pipeline of LMAC-TD: The time domain input signal $x(t)$ is fed into embedding model to obtain the classifier representations $h$. These classifier representations are then fed into the UNet Decoder which consists of series of conv-transpose operations and skip connections. The UNet Decoder produces a latent representation that is of the same shape as the output of the SepFormer Encoder. This representation $H_d$ is then combined with the encoder output $H_e$ and fed into the SepFormer MaskNet. The output of MaskNet $M$ is then element-wise multiplied and fed into the SepFormer Decoder to produce the time domain interpretation $i(t)$. During training, this time domain interpretation is then fed-back into the classifier to calculate the masking loss. {Light-blue boxes represent modules whose parameters are updated.}}
    \label{fig:pipeline}
    \vspace{-0.5cm}
\end{figure*}

\section{Methodology}

We describe the pipeline of LMAC-TD in Figure \ref{fig:pipeline}. The goal of this model is to produce a time-domain interpretation signal $i(t)$ for a time-domain input signal $x(t)$ that goes through the classifier $f(\cdot)$, composed of an embedding model $\text{Emb}(\cdot)$ and a classification head (OutHead in the diagram). We enhance the L-MAC architecture by adding the SepFormer's MaskNet, Encoder, and Decoder on top of the UNet Decoder.

First of all, the classifier representations are obtained by passing the input signal $x \in \mathbb R^T$ {first} through an input transformation $\text{InputTf}(\cdot)$ (e.g. STFT, Mel Spectrogram extraction), and {then through} the embedding model of the pre-trained classifier, $\text{Emb}(\cdot)$. We obtain a set of classifier representations such that, 
\begin{align}
    h = \text{Emb}(\text{InputTf}(x)).
\end{align}
{For our experiments, $\text{InputTf}(\cdot)$ represents the Mel-spectrogram computation, while $\text{Emb}(\cdot)$ refers to a CNN14 encoder \cite{kong2020panns}.} 
For $h$, we use the last 4 representations of the CNN14 encoder. The representations $h$ are then fed into a UNet decoder $U(\cdot)$, which is functionally similar to the decoder architecture in the L-MAC paper \cite{paissan24lmac}, but with adjusted strides and kernel sizes. We obtain the representation $H_d \in \mathbb R^{K \times T'}$ such that,
\begin{align}
    H_d = U(h).
\end{align}
In order to take into account contributions from both classifier representations and the encoded input audio directly, we combine $H_d$ with the output of the SepFormer encoder \cite{subakan2021attention} $H_e \in \mathbb R^{K\times T'}$. The convex combination of $H_e$ and $H_d$ is then fed into the SepFormer MaskNet to obtain the mask $M \in \mathbb R^{K \times T'}$ such that,
\begin{align}
    M = \text{MaskNet}\Big (\alpha H_d + (1-\alpha) H_e \Big ), 
\end{align}
with $\alpha \in [0, 1]$. We observe that calculating this combination through alpha allows the model to reach better faithfulness-audio quality trade-off. Finally, the mask $M$ is element wise multiplied with the Sepformer Encoder output $H_e$, and fed into SepFormer Decoder to obtain the time domain interpretation $i \in \mathbb R^T$ such that,
\begin{align}
 i = \text{SepFormerDecoder}\Big (M\odot H_e \Big).
\end{align}
The loss function is then calculated by comparing the classification results for the input signal $x$, the interpretation $i$, and the mask out version of the interpretation signal $i_\text{out}$, which is defined as,
\begin{align}
 i_\text{out} = \text{SepFormerDecoder}\Big ((1-M)\odot H_e \Big).
\end{align}
All in all, the training loss function is calculated following the L-MAC approach, 
\begin{align}
       \min_\theta & \lambda_\text{in} d(f(\text{InputTf}(x)) \| f(\text{InputTf}(i)) ) \\ & - \lambda_\text{out}  d(f(\text{InputTf}(x)) \| f(\text{InputTf}(i_\text{out})) ) + \lambda_\text{reg} R(i), \notag
\end{align}

where $d(\cdot||\cdot)$ is a divergence measure which measures the discrepancy between the classifier output $f(\text{InputTf}(x))$, and the classification of the interpretation $f(\text{InputTf}(i))$. In our experiments we use Cross-Entropy as $d(\cdot||\cdot)$. The second term tries to make sure that the masked-out interpretation signal $i_\text{out}$ minimizes its similarity to the input signal $x$ in terms of the classifier output. The coefficients $\lambda_\text{in}$ and $\lambda_\text{out}$ are used to control the relative strengths of the aforementioned mask-in and mask-out loss terms. Finally, the last term $R(i)$ is a regularization term that avoids trivial solutions. 
{In our experiments, we applied an $\ell^1$ penalty to the magnitude-STFT representation derived from the interpretation signal} such that, 
\begin{align}
    R(i) = \| \text{STFT}(i) \|_1.  
\end{align}

\section{Experiments}
In Sec.~\ref{sec:setup} we detail the experimental setup used to validate LMAC-TD.
We performed a quantitative analysis to establish the faithfulness of LMAC-TD-generated explanations, which we describe in Sec.~\ref{sec:quant}. Additionally, we conducted a user study to evaluate the user preference and present the results in Sec.~\ref{sec:qual}. 

\begin{table*}[t]
\caption{In-domain quantitative evaluation on the ESC50 dataset. Our results reveal that LMAC-TD achieves higher faithfulness scores (AI, AG, FF, Fid-In, SPS) compared to other methods.}
\label{tab:ID-ESC50}
\centering
\begin{tabular}{l|ccccccc}
\toprule
\textbf{Metric} & AI ($\uparrow$) & AD ($\downarrow$) & AG ($\uparrow$) & FF ($\uparrow$) & Fid-In ($\uparrow$) & SPS ($\uparrow$) & COMP ($\downarrow$) \\
\midrule
 {Saliency} & 0.00 & 15.79 & 0.00 & 0.05	& 0.07 & 0.39	& 5.48   \\ 
 {Smoothgrad} & 0.00	&15.71 & 0.00 & 0.03 &0.05	&0.42	&5.32 \\
 IG & 0.25 & 15.45 & 0.01 &0.07 & 0.13& 0.43 & 5.11 \\
 GradCAM & 8.50 & 10.11 & 1.47 &	0.17 &	0.33 &	0.34 &	5.64 \\
 Guided GradCAM & 0.00 & 15.61 &0.00& 0.05 &	0.06 &	0.44 & 5.12 \\
 Guided Backprop & 0.00	& 15.66 &0.00&0.05 & 0.06	 & 0.39 &	5.47 \\
 L2I, RT=0.2	& 1.63	&12.78 &	0.42	&0.11	&0.15&	0.25	&5.50 \\
{SHAP} & 0.00	&15.79 &	0.00 &	0.05 &	0.06 &	0.43 & 5.24 \\
L-MAC   & 36.25 &	\textbf{1.15}	& 23.50	&0.20 &	0.42	& 0.47 &	\textbf{4.71} \\ 
{L-MAC, FT, $\lambda_g=4$}  & 32.37 &	1.98	&18.74	& 0.21 &	0.41 &	0.43	& 5.20 \\
\midrule
\textbf{LMAC-TD}, $\alpha=1.00$ (ours)	& 66.00 &2.62  & 22.39 & \textbf{0.42} & 0.87 &	\textbf{0.86} &	 10.50 \\ 
\textbf{LMAC-TD}, $\alpha=0.75$ (ours)	& \textbf{69.75} & 2.10  & \textbf{28.07} & \textbf{0.42} & \textbf{0.91}&	\textbf{0.86} &	 10.53 \\ 
\textbf{LMAC-TD}, $\alpha=0.00$ (ours)	& 46.50 & 5.55  & 11.86 & \textbf{0.42} & 0.86 &	0.80 &	 10.88 \\ 

\bottomrule
\end{tabular}
\vspace{-0.3cm}
\end{table*}

\subsection{Setup}
\label{sec:setup}

\noindent \textbf{Metrics.} We use several quantitative metrics to compare LMAC-TD with state-of-the-art approaches. Specifically, we adopt the evaluation metrics employed in the original L-MAC paper \cite{paissan24lmac}. We use Average Increase (AI), which measures the increase in the classifier’s confidence for the interpretation with respect to the input sample, and Average Decrease (AD), which measures the confidence drop when removing the interpretation from the input (i.e. feeding $i_\text{out}$), and Average Gain (AG), similar to AI. Beyond these metrics, we use the Faithfulness (FF) metric defined in the L2I paper \cite{l2i} and the input fidelity (Fid-In) metric defined in the PIQ paper \cite{paissan2023posthoc}. We also use the Sparseness (SPS) \cite{chalasani2020concise} and Complexity (COMP) \cite{bhatt2020evaluating} metrics to evaluate the conciseness of the explanations. Due to space constraints, we invite the reader to refer to the respective papers for an in-depth presentation of the metrics. \\
\noindent \textbf{Implementation.} For all experiments, we use a CNN14 classifier \cite{kong2020panns} pre-trained on the VGGSound dataset \cite{Chen2020vggsound}. This classifier is then fine-tuned on the ESC50 dataset \cite{piczak2015dataset} with WHAM! noise \cite{Wichern2019WHAM} augmentation to mimic real-world conditions. It is trained on folds 1, 2, and 3, achieving 75$\%$ and 78$\%$ accuracy on folds 5 and 4, respectively. LMAC-TD is later trained on ESC50 with WHAM! noise augmentation. We refer to this data configuration when evaluating the explanations in the In-Domain setting. The UNet decoder employs a series of transposed convolution layers to align the last four latent representations with the encoder output $H_e$ shape. We benchmarked LMAC-TD with different $\alpha$ values to assess their impact. The remaining hyper-parameters are set as follows: $\lambda_\text{in} = 5$, $\lambda_\text{out}= 0.2$, and $\lambda_\text{reg}= 6$. We selected these hyperparameter combination based on our own qualitative evaluation of the produced audio quality, taking into account the faithfulness of the explanations.
Our implementation is based on the L-MAC implementation present in SpeechBrain 1.0~\cite{speechbrainV1} and is available through our companion website\footnote{\href{https://francescopaissan.it/lmac-td}{https://francescopaissan.it/lmac-td}}.\\
\noindent \textbf{Evaluation Data.} We conduct experiments to evaluate explanation faithfulness on In-Domain (ID) and Out-of-Domain (OOD) data. For the OOD setting, following L-MAC's procedure, we create mixtures of ESC-50 samples with the dominant class at $\SI{5}{\dB}$ SNR. We generate mixtures from folds 4 and 5. We also test LMAC-TD on audio contaminated with white noise and speech, using 3dB SNR mixtures and samples from the LJSpeech dataset~\cite{ljspeech17}. \\

\begin{table*}[t]
\caption{Out-of-Domain Quantitative Evaluation for the ESC50 Dataset. In out-of-distribution conditions, LMAC-TD achieves results comparable to L-MAC.}
\label{tab:OOD-ESC50}
\centering
\resizebox{.77\textwidth}{!}{
\begin{tabular}{l|ccccccc}
\toprule
\textbf{Metric} & AI ($\uparrow$) & AD ($\downarrow$) & AG ($\uparrow$) & FF ($\uparrow$) & Fid-In ($\uparrow$) & SPS ($\uparrow$) & COMP ($\downarrow$) \\
\midrule
 Saliency 	& 0.62 &	31.73	&0.07 & 0.06 &	0.12 & 0.76 & 11.06 \\
 Smoothgrad &0.12 &31.84 & 0.00 &0.06 &0.13 & 0.83 & 10.66 \\ 
  IG  & 0.37  & 31.15 &	0.03 &	0.12 & 0.26 &	0.87 & 10.22  \\
 L2I	&5.00	&25.65	&1.00 &	0.20 &	0.35 & 0.52 &	10.99 \\
 GradCAM &	14.12	& 17.62 & 7.46 & 0.25	& 0.00 & 0.91 & 9.66 \\
Guided GradCAM & 0.00	& 31.74 & 0.00 & 0.07	& 0.11	& 0.89 & 10.24 \\
Guided Backprop & 0.63 & 31.73 & 0.07 & 0.06 & 0.11 & 0.76 & 11.06\\
SHAP & 0.00 & 31.81 &	0.00 & 0.07 &	0.14 & 0.84 & 10.58 \\
 L-MAC  &\textbf{60.63} & 4.82 & \textbf{35.85} & 0.39 &	0.81 &	\textbf{0.94} & \textbf{9.61} \\
 L-MAC FT, $\lambda_g=4$  & 50.75 &6.73 & 26.00 & 0.39 & 0.78 &	0.84 &10.51 \\

\midrule
\textbf{LMAC-TD}, $\alpha=1.00$ (ours)	& 56.75 & 3.62 & 16.84 & \textbf{0.42} & \textbf{0.88} &	0.89 &	10.36 \\ 
\textbf{LMAC-TD}, $\alpha=0.75$ (ours)	& 59.50 & \textbf{3.42}  & 21.22 & 0.41 & 0.87 & 0.88	 &	10.35 \\ 
\textbf{LMAC-TD}, $\alpha=0.00$ (ours)	& 39.88 & 7.60  & 9.30 & 0.42 & 0.82 & 0.83 &	10.69 \\ 
\bottomrule
\end{tabular}
}
\vspace{-0.4cm}

\end{table*}

\subsection{Faithfulness Evaluation}
\label{sec:quant}
A very critical aspect of evaluating LMAC-TD explanations is establishing that they faithfully follow the classifier, meaning they effectively highlight portions of the input relevant for the classifier. We compare LMAC-TD with several other saliency-based methods in the literature. Namely, we compared with gradient-based methods such as saliency maps \cite{simonyan2014deep}, GradCAM \cite{gradcam}, SmoothGrad \cite{smilkov2017smoothgrad}, IntegratedGradients \cite{sundararajan2017axiomatic}, GuidedBackProp \cite{springenberg2015striving}, the decoder-based audio specific explanation method Listen-to-Interpret (L2I) \cite{l2i}, and we also include SHAP \cite{lundberg2017unified}.

In Table \ref{tab:ID-ESC50}, we report the quantitative results obtained for the ID experiments with three different LMAC-TD configurations. 
We observe that, with $\alpha = 0.75$, LMAC-TD outperforms standard L-MAC in all faithfulness metrics except AD, which remains comparable. We note that we have calculated the COMP metric (a complexity metric) on the magnitude-STFT representations of the explanation signal $i$ which is produced in the time domain. This tends to give large COMP metrics for LMAC-TD implying that LMAC-TD explanations are complex (high entropy - uniformly distributed in the STFT domain). However, LMAC-TD explanations remain understandable, and the explanations are still preferable for the users over the alternatives as we show in Section \ref{sec:qual} with a user study.


 We note that the results show that lowering $\alpha$ reduces the faithfulness metrics. This makes sense since larger $\alpha$ more strongly incorporates the classifier representations, increasing the classifier influence on the explanations. 
 

\begin{table*}[t]
\caption{Additional results obtained on the ESC50 dataset with White Noise and LJSpeech contamination. 
}
\label{tab:ESC50extra}

\centering
\resizebox{.77\textwidth}{!}{
\begin{tabular}{l|ccccccc}
\toprule
\textbf{Metric} & AI ($\uparrow$) & AD ($\downarrow$) & AG ($\uparrow$) & FF ($\uparrow$) & Fid-In ($\uparrow$) & SPS ($\uparrow$) & COMP ($\downarrow$) \\
\midrule
 & \multicolumn{7}{c}{\textit{Classification on ESC50, White Noise Contamination, 38.6\% accuracy}} \\
L2I @ 0.2 &	0.00 &	19.41 &	0.21 &	0.11 &	0.04 	&36.62& 	\textbf{7.32} \\
L-MAC  &	\textbf{83.62} & \textbf{1.50} & \textbf{56.12} &	0.33 &	\textbf{0.86} &	\textbf{0.92} &	{10.03}  \\
\midrule
\textbf{LMAC-TD}, $\alpha=1$ (ours)	 &  55.00  & 4.32 & 16.15 &  0.35 &	0.81 &	0.82 & 10.75 
\\ 
\textbf{LMAC-TD}, $\alpha=0.75$ (ours)	& 63.50 & 3.06 & 24.57 & 0.35 & 0.83 &	0.84 &	10.71  \\ 
\textbf{LMAC-TD}, $\alpha=0.00$ (ours)	& 33.50 & 10.03 & 6.83  & \textbf{0.36} & 0.71  &0.75	 &	11.09  \\ 

\midrule
 & \multicolumn{7}{c}{\textit{Classification on ESC50, LJSpeech Contamination, 79.3\% accuracy}} \\
L2I @ 0.2 & 	1.75 &	29.49 &	0.27 & 	0.15 & 	0.18 & 	0.79 & 	\textbf{9.56} \\
L-MAC &	\textbf{70.75} & \textbf{2.73} & \textbf{39.64} & 0.33 & 0.83 	& \textbf{0.93} &	9.70 \\
\midrule
\textbf{LMAC-TD}, $\alpha=1.00$ (ours)	& 59.13 & 3.05 & 17.18 & 0.43 & \textbf{0.90} &	0.88 &	10.33  \\ 
\textbf{LMAC-TD}, $\alpha=0.75$ (ours)	& 62.63 & 2.95 & 21.57 & 0.43 & \textbf{0.90}  &	0.88 &	10.33  \\ 
\textbf{LMAC-TD}, $\alpha=0.00$ (ours)	& 44.25 & 6.50 & 9.88 & \textbf{0.44} & 0.87  &	0.83 &	10.69  \\ 

\bottomrule
\end{tabular}
}
\end{table*}

In Table \ref{tab:OOD-ESC50}, we compare the results obtained on audio mixtures created with ESC50 data.  
Unlike the ID case, LMAC-TD with $\alpha = 0.75$ outperforms both L-MAC and STFT-masking approaches in AD, FF, and Fid-In. With regards to the effect of $\alpha$, we observe that in general the faithful metrics are better as expected with larger $\alpha$, however in terms of FF $\alpha=0$ remains comparable. 
These results suggest that the explanations generated with LMAC-TD are well-aligned with the classifier in this OOD setting as well.

Finally, Table \ref{tab:ESC50extra} presents the results on the ESC50 dataset with white noise and LJSpeech contamination\footnote{In the table, we have included only the top two methods for comparison; the complete set of results is available on the companion website.} to explore more OOD settings. 
In these settings, LMAC-TD achieves the best overall faithfulness scores with $\alpha = 0.75$. Compared to standard L-MAC we observe a drop in terms of the performance in some metrics such as AI, AD AG, however in terms of FF and Fid-In LMAC-TD achieves better scores. 
We also would like to note that, similar to the ID setting, in all OOD settings larger $\alpha$ typically results in better faithfulness scores. 





\subsection{Subjective Evaluation}

\begin{figure}[h!]
    \centering
    \includegraphics[width=0.99\linewidth]{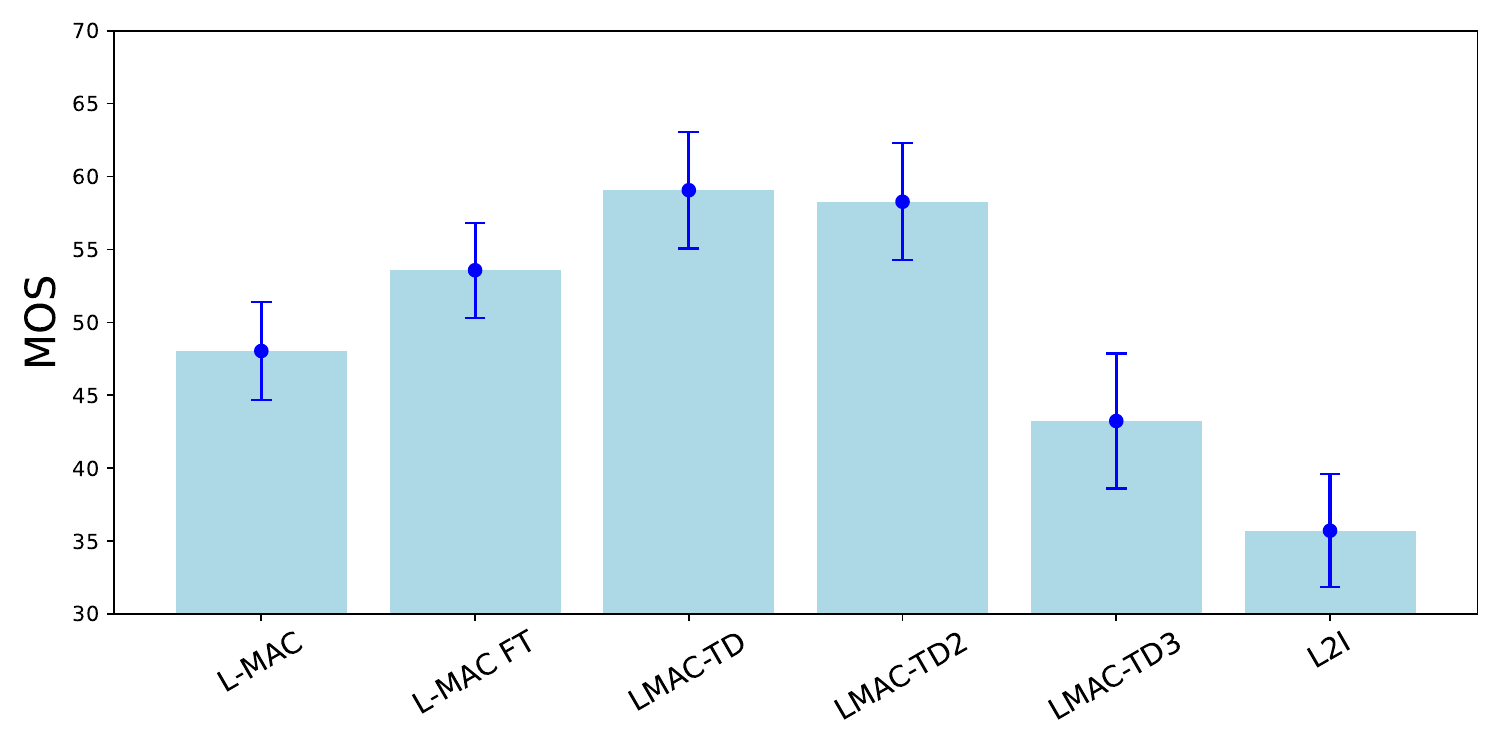}
    \vspace{-.5cm}
    \caption{Mean-Opinion Score (MOS) for the interpretations. Confidence intervals at $0.95$ are reported as error bars (in blue). In the Figure, LMAC-TD corresponds to the configuration with $\alpha=1$, LMAC-TD2 to the configuration with $\alpha=0.75$, and LMAC-TD3 to the configuration with $\alpha=0$.}
    \label{fig:ustudy}
\end{figure}

\label{sec:qual}
To determine the perceived quality of the generated interpretations we conducted a user study with 19 participants using the WebMushra interface \cite{webmushra}, following the evaluation pipeline used for L-MAC \cite{paissan24lmac}. We presented nine audio samples to the users and we asked the users to assign a score (1 to 100) based on (i) how well the interpretation corresponds to the part of the input audio associated with the predicted class and (ii) the perceived audio quality. We present the outcome of the user study in Fig.~\ref{fig:ustudy} with the confidence intervals at $0.95$. LMAC-TD outperforms the other methods when $\alpha \in \{1, 0.75\}$. With $\alpha=0$, LMAC-TD obtains a Mean-Opinion Score (MOS) comparable to the original L-MAC, while still outperforming L2I. 

We also observe that LMAC-TD is preferred over LMAC-FT (the fine tuned version of LMAC \cite{paissan24lmac}). 
We note that LMAC-TD simplifies the pipeline by removing the need for interpreter finetuning, while improving perceived audio quality over LMAC-FT. The audio samples from the user study are accessible through our companion website\textsuperscript{1} together with the per-sample MOS.

In summary, faithfulness and qualitative evaluations demonstrate that LMAC-TD with $\alpha = 1$ and $\alpha = 0.75$ both offer better subjective audio quality compared to the baselines, while not compromising significantly from faithfulness. 

\section{Conclusions}

This paper presents a novel approach, called LMAC-TD, to generate listenable explanations directly in the time domain. LMAC-TD builds on top of recent literature in interpretability and source separation to enhance the audio quality of the explanation. We propose to include a time-domain source-separation architecture to decode listenable explanations from the classifier representations. Through an extensive experimental analysis, we conclude that LMAC-TD outperforms the other methods in terms of perceived audio quality and achieves better or comparable results in faithfulness metrics.

\bibliographystyle{IEEEtran}
\bibliography{refs}

\end{document}